\documentclass[graybox]{svmult}

\usepackage{mathptmx}       
\usepackage{helvet}         
\usepackage{courier}        
\usepackage{type1cm}        
\usepackage{makeidx}         
\usepackage{graphicx}        
\usepackage{multicol}        
\usepackage[bottom]{footmisc}

\makeindex             

\begin{document}

\title*{Pulsating stars harbouring planets}
\author{A. Moya}
\institute{Departamento de Astrof\'{\i}sica, Centro de
     Astrobiolog\'ia (INTA-CSIC), PO BOX 78, 28691 Villanueva de la
     Ca\~nada, Madrid, Spain, \email{amoya@cab.inta-csic.es}}
%
%
\maketitle

\abstract*{Why bother with asteroseismology while studying exoplanets?
  There are several answers to this question. Asteroseismology and
  exoplanetary sciences have much in common and the synergy between
  the two opens up new aspects in both fields. These fields and
  stellar activity, when taken together, allow maximum extraction of
  information from exoplanet space missions. Asteroseismology of the
  host star has already proved its value in a number of exoplanet
  systems by its unprecedented precision in determining stellar
  parameters. In addition, asteroseismology allows the possibility of
  discovering new exoplanets through time delay studies. The study of
  the interaction between exoplanets and their host stars opens new
  windows on various physical processes. In this review I will
  summarize past and current research in exoplanet asteroseismology
  and explore some guidelines for the future.}

\abstract{Why bother with asteroseismology while studying exoplanets?
  There are several answers to this question. Asteroseismology and
  exoplanetary sciences have much in common and the synergy between
  the two opens up new aspects in both fields. These fields and
  stellar activity, when taken together, allow maximum extraction of
  information from exoplanet space missions. Asteroseismology of the
  host star has already proved its value in a number of exoplanet
  systems by its unprecedented precision in determining stellar
  parameters. In addition, asteroseismology allows the possibility of
  discovering new exoplanets through time delay studies. The study of
  the interaction between exoplanets and their host stars opens new
  windows on various physical processes. In this review I will
  summarize past and current research in exoplanet asteroseismology
  and explore some guidelines for the future.}

\section{Introduction}
\label{sec:1}

In a review devoted to the field of study of pulsating stars
harbouring planets, S. Vauclair \cite{reviewsilvie} wondered ``why
bother with asteroseismology in exoplanet research?''. She gave
several answers to this question. This review is based on further
elaboration of these topics. In addition, I will discuss recent
developments which relate to this question.
  
The potential of asteroseismology for the study of exoplanetary
systems was already evident soon after the discovery of the first exoplanets. 
However, it is only in the last six years that the interaction between
exoplanet research and asteroseismology began to show promising results.
Let us begin by investigating some of the topics that resulted from
Vauclair's question.

\section{Synergy: (almost) the same instruments are used}

Asteroseismology and exoplanet detection are both based on the
analysis of time series. Currently, the best way of detecting
exoplanets is through radial velocity measurements (ground based
observations) or the transit technique which demands a high
photometric precision, mainly attained from space
  missions. Asteroseismology has the same requirements.  Most of this
review is devoted to the contribution of asteroseismology to exoplanet
research. The contribution of exoplanet science to asteroseismology
cannot be neglected since exoplanets may offer an explanation for some
asteroseismological puzzles.

The synergy between the two fields will always co-exist in present and
future exoplanet space missions. It is in the characterization of the 
host star that asteroseismology plays a key role 
\cite{corot,kepler,plato}.

\section{Noise: one person's noise is another person's signal}

Historically, the exoplanet community has treated the variability of
the host star, whether activity or pulsation, as noise which
difficults the detection of exoplanets \cite{dumusque}. Attempts have
been made to reduce this ``noise'' using different techniques.
However, sometimes the noise is a signal that can assist in the
detection of the exoplanet and characterization of the central star.
This is the idea behind the Plato Data Center, where light curves will
be stored and studied by a collaboration between the exoplanet and
asteroseismological communities.

\section{Characterization: Obtain precise values of the parameters of 
exoplanets-host stars}

The study and characterization of stars hosting planets is an 
essential requirement in understanding the physics of exoplanets. This can
be illustrated by a summary of the most successful techniques that are
presently available for the detection of exoplanets.  Currently, five main
techniques are in use, but by far the most common are radial velocities,
transits and direct imaging.

\begin{itemize}

\item Radial velocities.  This technique uses the periodic variation in
  the radial velocity due to the gravitational perturbation of the orbiting
  planet. This method provides a lower limit on the mass ratio.

\item Transits.  An orbiting planet crossing the line of sight results
  in a shallow dip in the stellar light curve.  Analysis of this 
  eclipse gives the orbital parameters and ratio of the radii.

\item Direct imaging. It is the direct detection of the planet in an
  image. The observational data is the planet luminosity. Measuring
  the luminosity of the planet and modelling the stellar age allows
  the mass of the planet to be determined.

\end{itemize}

It is clear that uncertainty in the general physical characteristics
of the central star leads directly to uncertainty in the physical
characteristics of the exoplanet \cite{sowth}.  There are a number of
papers which discuss the use of asteroseismology in precise
characterization of stellar parameters
\cite{hans,stello1,stello2,basu,quirion,gai,mulet}.  Most of these
studies are for single main-sequence stars similar to the Sun.

All these studies stress the importance of asteroseismology.  For solar-like
stars the radius can be estimated with an uncertainty of about 3 percent,
the mass to about 5 percent and the age to 5--10 percent.  However, accuracy 
is not the same as precision; these uncertainties depend entirely on our 
understanding of stellar structure and evolution of the particular star.

Have we achieved the predicted precision in these parameters?  A
number of studies concerned with modeling of the observational data
are available \cite{chap,metc}.  These studies show that the stellar
radius is estimated to better than 1~percent for main-sequence
solar-like stars.  The mass is estimated with a precision of
about 5~percent and the mean density to $<0.2$ percent. Currently,
there are fifteen papers in which pulsating stars with planets
  are studied:
\cite{muarae1,iota1,46,hatp7,52,wasp33,hr8799_1,beta,17,99,v391,idrac}.
Most of these studies concern solar-like stars or post-MS stars with
solar-like pulsations. Not all these papers discuss the pulsational
characteristics of the host star in the same detail.  Some highlights
of these studies are itemized below.

\begin{itemize}

\item $\mu$ Arae.  This was the first star where asteroseismology was
   applied to exoplanet research.  The system consists of a giant 
   exoplanet orbiting a solar-like star \cite{muarae1,muarae2,muarae3,muarae4}.

\item $\iota$ Horologii.  This is another giant exoplanet orbiting a
  solar-like star \cite{iota1,iota2,iota3}. The stellar age and He
  content estimated from asteroseismic analysis indicates membership
  of the Hyades cluster. This result led to a revision of the stellar
  mass and therefore to the mass of the planet.

\item KIC~9904059. Asteroseismology of this star shows that a suspected
  planet is in fact a companion star \cite{99}.

\item HAT-P-7 \cite{hatp7}, HD 46375 \cite{46}, HD17156 \cite{17}.
  In these studies, stellar characterization using asteroseismology
  attained the predicted precision and confirmed the parameters found
  in previous studies.

\item HR\,8799. This is the only pulsating star with a planetary
  system that has been directly imaged \cite{hr8799_2}. The age of the
  system and the mass of the central star are the key parameters in
  this case. HR\,8799 is a main-sequence $\lambda$ Bootis star, which
  renders standard techniques of age determination not very
  reliable. The pulsations in HR\,8799 offer unique
  opportunities for further study of this interesting system.

\end{itemize}

It should be borne in mind that all these characterizations are very model 
dependent \cite{lebreton}.  A better understanding of stellar physics 
is required to fully assess these predictions.

\section{Planet detection using the timing method}

Asteroseismology can be also used to discover planets using the
timing method \cite{silvotti} which requires an accurate astronomical 
clock. The general requirements for stellar pulsation modes to be 
used in this way are 1) very stable pulsational periods and amplitudes,
2) several independent high amplitude modes, and 3) the periods must be
short. Some types of pulsating stars that may fulfill these requirements 
are SdB pulsators, roAp and $\delta$ Scuti stars.  For example,
a planet has been found in the SdB star V391 Peg by this method \cite{v391}.  
The {\it EXOTIME} international consortium is devoted to the use of
the timing method for detecting planets around SdB stars \cite{schuh,lutz}.

\section{Interaction between star and planet}

In addition, there are a number of on-going projects concerning star -
planet interactions.  For example, we know that most of the exoplanets
so far discovered have metal-rich host stars, but why this is the case
is not known \cite{sousa}.  We do not know whether the over-abundance
of metals is intrinsic to the star or a result of accretion
\cite{muarae1}.  Stars with planets seem to have a lower Li abundance
than stars without planets \cite{israelian}.  It is possible that Li
depletion may be due to angular momentum transport and additional
mixing caused by the interaction between the star and planet(s)
\cite{eggenberger}.  Another possibility is an accretion process
\cite{theado}.  Asteroseismology may well shed light on these
problems.

An asteroseismic study of HR\,8799 has shown, for the first time, a
misalignment between the rotation axis of the host star and the orbital
plane of the debris disk.  In addition, there is misalignment with the 
orbital plane of the three known planets \cite{wright}.

Finally, as already demonstrated for KIC~9904059, asteroseismology can 
assist in uncovering false detections of exoplanets \cite{tingley}.

\section{Future prospects}

The European Space Agency, in its ``Cosmic Vision'' plan, proposes
four main objectives which are presented as questions.  The first 
question is: what are the conditions for planet formation and the 
emergence of life?  To answer this question, an advisory team
suggested the following three steps: 1) discovering exoplanets, 2) 
characterization of these systems, and 3) the search for bio-markers
\cite{roadmap}. As I have described in this review, asteroseismology
is an important tool in planet detection and is essential for
characterization of the system.

Future efforts will concentrate in the accurate characterization of host 
stars, especially for the brighter stars.  Attempts should be made to 
discover more planetary systems around other types of pulsating stars and
the use of asteroseismology to recognise false detections.  The study of the 
interaction between star, planets and debris disk should be intensified.  
Finally, we need to apply asteroseismology to systems with exoplanets 
around very low mass stars and brown dwarfs.

To facilitate these objectives, there are a number of current and
  future ground-based instruments (HARPS, HARPS-N, CARMENES,
ESPRESSO, etc.) which have been specifically designed for exoplanet
studies and may also be used for asteroseismology.  Space missions
such as {\it MOST, CoRoT, Kepler}, and {\it BRITE} have already
demonstrated the extraordinary synergy between exoplanet research and
asteroseismology. Three additional space missions directed towards
exoplanet/asteroseismology are currently proposed ({\it Plato, EcHO,
  TESS}), but only some of these are likely to be accepted.  There is
no doubt that further space missions are required to answer the
questions listed above.

\begin{acknowledgement}
The   author   acknowledges    the   funding   of   AstroMadrid   (CAM
S2009/ESP-1496). This  research has been funded by  the Spanish grants
ESP2007-65475-C02-02, AYA 2010-21161-C02-02.
\end{acknowledgement}
\bibliographystyle{spmpsci.bst}

\end{document}